\documentclass[reprint,aps,pre,amsmath,amssymb,superscriptaddress]{revtex4-1}
\usepackage{graphicx,hyperref}
\hypersetup{colorlinks=true, linkcolor=blue, citecolor=blue, urlcolor=blue}

\renewcommand{\l}[0]{\left}
\renewcommand{\r}[0]{\right}
\let\f=\frac
\newcommand{\fp}[2]{\l( \f{#1}{#2} \r)}

\renewcommand{\v}[1]{\mathbf{#1}}

\renewcommand{\d}[0]{\ensuremath{\operatorname{d}\!}}
 
\let\del=\partial
\newcommand{\pd}[2]{\f{\del #1}{\del #2}} 
\newcommand{\Ang}[0]{\, \mathring{\mathrm{A}}}

\newcommand{\kB}[0]{k_\t{B}}
\newcommand{\K}[0]{\, \t{K}}
\newcommand{\e}[0]{\mathrm{e}}
\newcommand{\avg}[1]{\langle #1 \rangle}
\newcommand{\Avg}[1]{\l< #1 \r>}

\renewcommand{\t}[1]{\text{#1}}

\begin{document}

\title{Free energy of grain boundary phases: Atomistic calculations for $\Sigma5(310)[001]$ grain boundary in Cu}
\author{Rodrigo Freitas}
\altaffiliation[Current affiliation: ]{Department of Materials Science and Engineering, Stanford University, Stanford, CA 94305, USA}
\email{freitas@stanford.edu}
\affiliation{Department of Materials Science and Engineering, University of California, Berkeley, CA 94720, USA}
\affiliation{Lawrence Livermore National Laboratory, Livermore, CA 94550, USA}
\author{Robert E. Rudd}
\affiliation{Lawrence Livermore National Laboratory, Livermore, CA 94550, USA}
\author{Mark Asta}
\affiliation{Department of Materials Science and Engineering, University of California, Berkeley, CA 94720, USA}
\affiliation{Materials Sciences Division, Lawrence Berkeley National Laboratory, Berkeley, CA 94720, USA}
\author{Timofey Frolov}
\affiliation{Lawrence Livermore National Laboratory, Livermore, CA 94550, USA}
\date{\today}
\begin{abstract}
  Atomistic simulations are employed to demonstrate the existence of a well-defined thermodynamic phase transformation between grain boundary (GB) phases with different atomic structures. The free energy of different interface structures for an embedded-atom-method model of the $\Sigma 5 (310) [001]$ symmetric tilt boundary in elemental Cu is computed using the nonequilibrium Frenkel-Ladd thermodynamic integration method through molecular dynamics simulations. It is shown that the free-energy curves predict a temperature-induced first-order interfacial phase transition in the GB structure in agreement with computational studies of the same model system. Moreover, the role of vibrational entropy in the stabilization of the high-temperature GB phase is clarified. The calculated results are able to determine the GB phase stability at homologous temperatures less than $0.5$, a temperature range particularly important given the limitation of the methods available hitherto in modeling GB phase transitions at low temperatures. The calculation of GB free energies complements currently available $0\K$ GB structure search methods, making feasible the characterization of GB phase diagrams.
\end{abstract}
\maketitle

\section{Introduction}
  Properties of polycrystalline materials are greatly influenced by the presence of internal interfaces called grain boundaries (GB) \cite{Balluffi95}. Because of their importance the structural, energetic and kinetic properties of GBs have been extensively investigated, both experimentally and through computational modeling. Early atomistic simulation studies immediately recognized the inherent multiplicity of GB structures at $0\K$ for a fixed relative orientation of the neighboring bulk grains \cite{vitek}. More recently, an idea of GB phase behavior has surfaced and gained growing attention due to accumulating experimental evidence of discontinuous transitions in materials properties observed in both bicrystals and polycrystalline materials \cite{Cantwell20141,Divinski2012,PhysRevLett.59.2887,Dillon2016324,Dillon20075247}. Specifically, an unusual non-Arrhenius diffusion of Ag \cite{Divinski2012} and Au \cite{au} radioactive isotopes was measured in a well-characterized symmetric tilt $\Sigma 5(310)[001]$ Cu GB, indirectly suggesting a temperature-induced structural transformation. These experiments suggested that similar to bulk materials, GBs have different structures depending on the temperature, pressure, and chemical composition. However, experimental evidence of GB phases is often indirect and in most cases does not provide the atomic-scale structure of these phases especially at high temperature. 

  Theoretical investigation into the possibility of multiple phases of interfaces dates back to work of Gibbs, who considered finite variations in the state of an interface and predicted that if competing structures are possible the interface state with lowest free energy should be the most stable \cite{Gibbs}. More recently, thermodynamic analysis \cite{Hart:1972aa,Frolov:2015ab}, phase-field models \cite{Tang06,Tang06b}, lattice-gas models \cite{Rickman2016225,Rickman20161,Rickman201388}, and atomistic simulations \cite{Olmsted2011,Frolov2013,Frolov2013PRL,Frolov2016,PhysRevB.92.020103,zhu2018,frolov2018grain} were successfully used to study phases and phase transitions at GBs. While the thermodynamic analysis and analytical models formulated rigorous rules for GB phase coexistence, the approach of atomistic simulations is appealing because it promises to predict atomic structures and properties of GB phases. The modeling of GB phase transitions, or GB structures in general, using atomistic simulations faces the same challenges as structure prediction of bulk materials with an additional complication that the structure is confined between two misoriented crystals, which introduces new degrees of freedom. While the common simulation methodology known as the $\gamma$-surface methods attempts to generate GB structure from two perfect half crystals, studies have demonstrated that more advanced structure search algorithms are required to predict the ground state and metastable states \cite{duffy,tasker,duffy2,DUFFY84a,Phillpot1992,Phillpot1994,Frolov2013,Chua:2010uq}. Recently, a new computational approach for GB structure prediction has been proposed \cite{zhu2018}. The tool is based on the USPEX structure search code and uses evolutionary algorithms to perform a grand-canonical search of GB structure. This computational tool augmented with unsupervised machine-learning post-processing analysis identifies ground states as well as metastable GB phases at $0\K$ temperature. 

  While the structure search methods can thoroughly explore a diverse range of GB configurations, their predictions regarding the finite-temperature GBs may still be ambiguous. The algorithms minimize GB energy at $0\K$ and often yield multiple GB phases with close energies, making it difficult to predict which GB phase will be observed at finite temperature. It is clear that the $0\K$ analysis alone cannot predict GB phase diagrams and temperature-induced GB phase transitions.

  High-temperature molecular dynamics (MD) can simulate such transitions directly and in some cases can even identify transition temperature \cite{Olmsted2011,Frolov2013,Alfthan07}. However, these types of simulations are only effective in a relatively narrow range of temperatures. The transformation requires nucleation of a new GB phase and the nucleation barrier can be prohibitively large to observe the structural change on conventional MD time scales. As a result, conventional MD approaches cannot simulate such transformations below about half of the melting point. Therefore, following the thermodynamic analysis of Gibbs, the free-energy calculations of different GB phases may be a more effective modeling approach to predict finite-temperature GB phases and transition temperatures. Prior modeling work calculated free energy of individual GBs as a function of temperature and composition using methodologies that combined the harmonic approximation for vibrational free energy with the integration of the adsorption equation \cite{review_mark,Frolov2012b,Foiles2010,obrien,Foiles94,Broughton86a,245,247,250,extra}. In this work, we use the nonequilibrium Frenkel-Ladd (FL) method to calculate free energies of different GB phases of the same boundary and predict GB phase transitions in a model elemental metal system. 

  The remainder of this paper is organized as follows. In Sec.~\ref{sec:methodology} we describe the methodology of the GB free-energy calculations. Section \ref{sec:simulations} presents the atomistic simulation details, including system geometry and simulation setup for the free-energy calculations. In Sec.~\ref{sec:results} we present the results of the free-energy calculations for different GB phases and compare the accuracy of different methods. In Sec.~\ref{sec:conclusions} we summarize the main findings and present an outlook for applying the FL method in the calculation of free energies of other GBs.

\section{Methodology\label{sec:methodology}}
\begin{figure*}
    \centering
    \includegraphics[width=0.90\textwidth]{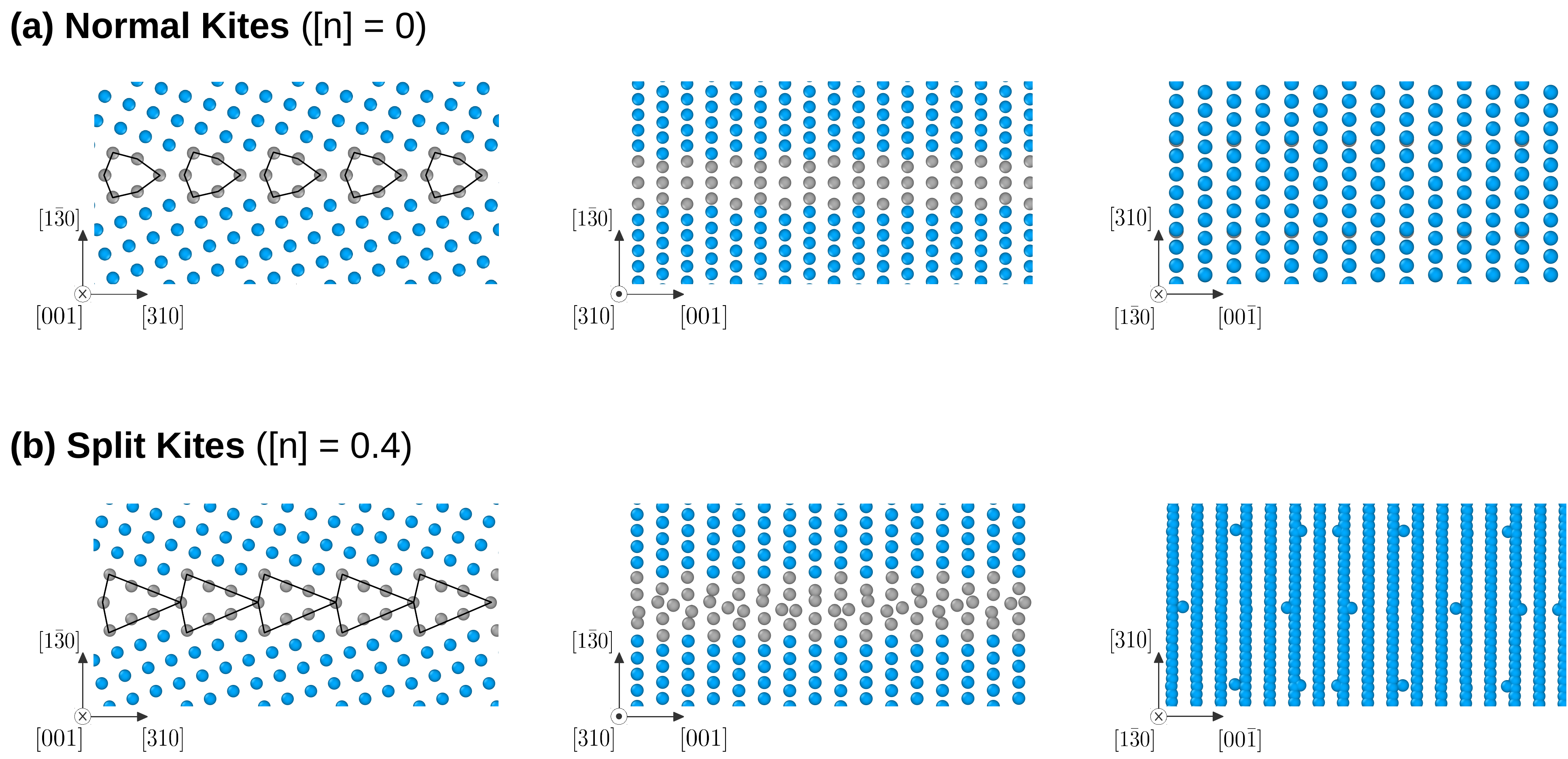} 
    \caption{\label{fig:structure}Different phases of the $\Sigma5(310)[001]$ GB in Cu \textbf{(a)} normal kites (NK) and \textbf{(b)} split kites (SK). Three different views of each structure are shown: in the left-hand side the tilt axis is normal to the plane of the figure, the middle panel shows side view with the tilt axis parallel to the plane of the figure, and the right-hand side panel shows the GB plane as viewed from the top. The periodic units of the SK and NK phases have different dimensions: the SK phase is a $10\times2$ reconstruction relative to the NK phase \cite{Frolov2013}.}
  \end{figure*}

\begin{figure}
    \centering
    \includegraphics[width=0.48\textwidth]{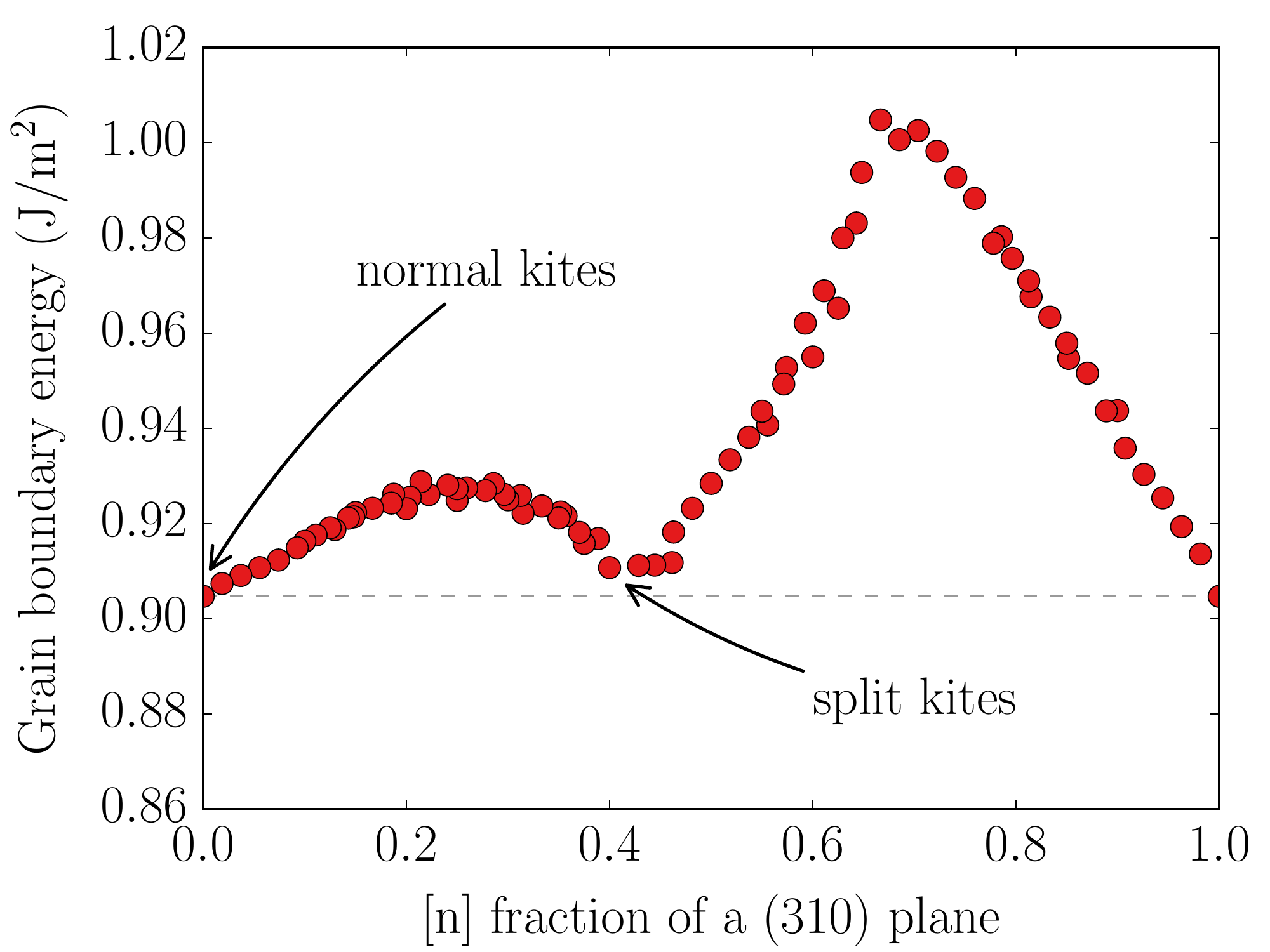}
    \caption{\label{fig:energy}Energy of the $\Sigma5(310)[001]$ GB as a function of number of atoms $[n]$ measured as a fraction of atoms in a $(310)$ bulk plane \cite{Frolov2013}. The two minima, at $[n] = 0$ and $0.4$ -- correspond to the NK and SK phases, respectively. The dotted line marks the energy of the normal kites $0.9048 \, \t{J/m}^2$, while the split kites have energy $0.9112 \, \t{J/m}^2$.}
  \end{figure}

  In this paper we study the $\Sigma 5(310)[001]$ symmetric tilt GB in Cu as modeled by an embedded-atom method (EAM) \cite{mishin_cu} potential. According to this potential, the boundary can exist in two different ordered phases called normal kites (NK) and split kites (SK) \cite{Frolov2013}, which are illustrated in Fig.~\ref{fig:structure}. These two structures correspond to two GB energy minima as a function of GB atomic density, $[n]$, as shown in Fig.~\ref{fig:energy}. The number of atoms $[n]$ was introduced in Ref.~\onlinecite{Frolov2013} and for this boundary it is defined as $\t{modulo}(N,N_{310})/N_{310}$, where $N$ is the total number of atoms in the bicrystal and $N_{310}$ is the number of atoms in a (310) plane in the simulation box in the bulk region. The NK phase is composed of kite-shaped structural units and has atomic density $[n] = 0$, while the SK phase has more complicated structure and has a higher density of $[n] = 0.4$. The difference in $[n]$ suggests that extra atoms equal to 0.4 fraction of (310) plane have to be inserted in NK structure to obtain SK. At $0\K$, the NK phase is the ground state with energy $\gamma_\t{nk}(0\K) = 0.9048 \, \t{J/m}^2$, while the SK phase has higher energy: $\gamma_\t{sk}(0\K) = 0.9112 \, \t{J/m}^2$. This energy difference is of only $0.7\%$, which is much smaller than the expected changes in GB free energy due to increasing temperature \cite{Frolov2012b,Foiles2010,obrien,Foiles94,Broughton86a}. Perhaps not surprisingly, MD simulations with the GB connected to an open surface demonstrated that at $800\K$ the NK phase transforms into the SK phase. The open surface supplies the extra atoms -- about $[n] = 0.4$ atomic fraction of a $(310)$ plane -- necessary for the transformation. While this methodology demonstrates that at $800\K$ the SK phase becomes more stable, it falls short of predicting the exact transition temperature because below $800\K$ the transition cannot be observed on the MD time scale due to prohibitively slow GB diffusion. Thus, this model system is ideal to study how free-energy calculations can predict GB phase transitions at relatively low temperatures.

  For the remainder of the paper when we refer to physical properties of these GB phases a subscript ``nk'' and ``sk'' will be used (e.g., $\gamma_\t{nk}$ and $\gamma_\t{sk}$). If there is no need for distinction we use the ``gb'' subscript instead (e.g., $\gamma_\t{gb}$).

  \subsection{Grain boundary free-energy calculations}
\begin{figure}
      \centering
      \includegraphics[width=0.48\textwidth]{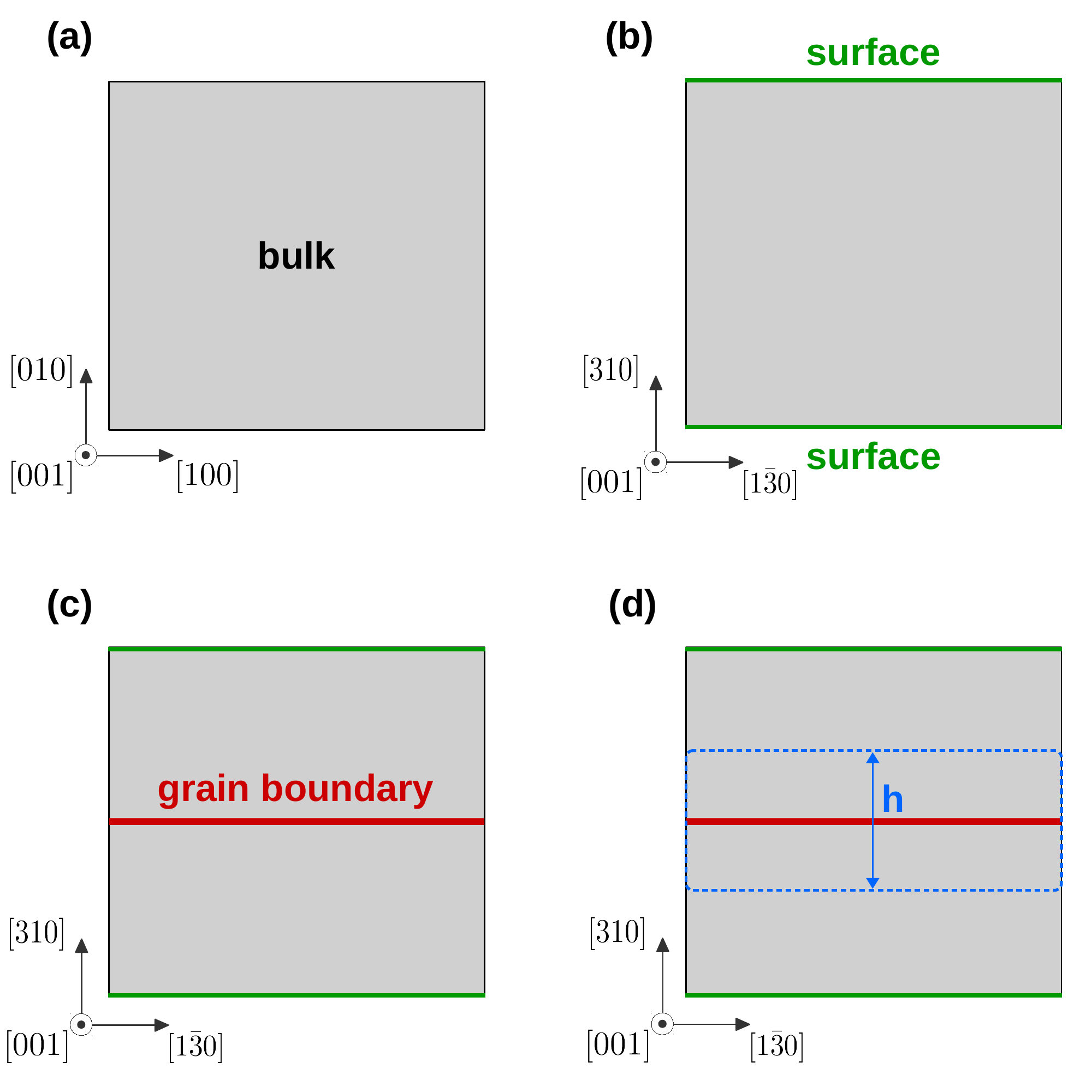}
      \caption{\label{fig:simulation_boxes}Schematics of the simulation blocks used for GB free-energy calculations. \textbf{(a)} Bulk system with full periodic boundary conditions applied. \textbf{(b)} Single crystal slab system with two free surfaces and same crystallographic orientation as the GB system. \textbf{(c)} Bicrystal system with the $\Sigma5(310)[001]$ GB and two free surfaces. \textbf{(d)} Bicrystal system partitioned into a subsystem region with the same cross section as the simulation box and height $h$ centered around the GB.}
    \end{figure}

    Consider a bicrystal with a symmetric GB and two surfaces illustrated in Fig.~\ref{fig:simulation_boxes}(c) (the case of asymmetric boundaries will be considered below). GB free energy per unit area, $\gamma_\t{gb}$, can be expressed as \cite{Gibbs}
    \begin{equation}
      \label{eq:gamma_gb}
      \gamma_\t{gb} A_\t{gb} = F_\t{gb} - \gamma_\t{surf} A_\t{surf} - N_\t{gb} f_\t{bulk}
    \end{equation}
    where $A_\t{gb}$ is the GB area, $F_\t{gb}$ is the total free energy of the bicrystal system, $A_\t{surf}$ is the total surface area ($A_\t{surf} = 2 A_\t{gb}$), and $N_\t{gb}$ is the total number of atoms in the bicrystal. The remaining terms in Eq.~\eqref{eq:gamma_gb} are $\gamma_\t{surf}$ and $f_\t{bulk}$ -- the surface free energy per unit area and the bulk free energy per atom, respectively -- defined as follows. Consider a perfect bulk system with full periodic boundary conditions and $N_\t{bulk}$ atoms, as shown in Fig.~\ref{fig:simulation_boxes}(a). We define 
    \begin{equation}
      \label{eq:f_bulk}
      f_\t{bulk} = F_\t{bulk} / N_\t{bulk},
    \end{equation}
    where $F_\t{bulk}$ is the total free energy of the bulk system in Fig.~\ref{fig:simulation_boxes}(a). Consider now the system in Fig.~\ref{fig:simulation_boxes}(b): this is a single crystal slab with the same crystallographic orientation and dimensions as the bicrystal system, but it does not contain a GB. If the total surface area of this system is $A_\t{surf}$ and the total free energy is $F_\t{surf}$, the surface free energy is:
    \begin{equation}
      \label{eq:gamma_surf}
      \gamma_\t{surf} A_\t{surf} = F_\t{surf} - N_\t{surf} f_\t{bulk},
    \end{equation}
    where $N_\t{surf}$ is the number of atoms in the system in Fig.~\ref{fig:simulation_boxes}(b).

    Hence, Eqs.~\eqref{eq:gamma_gb}, \eqref{eq:f_bulk}, and \eqref{eq:gamma_surf} can be used to determine $\gamma_\t{gb}$ for any bicrystal system, given that we can determine the absolute free energy of the system illustrated in Figs.~\ref{fig:simulation_boxes}(a)--(c), i.e., given that $F_\t{gb}$, $F_\t{surf}$, and $F_\t{bulk}$ can be computed. In the next two sections we present two methods, namely, the FL and the quasi-harmonic approximation (QHA), that enable the calculation of the absolute free energy of systems in Figs.~\ref{fig:simulation_boxes}(a)-(c) using atomistic simulation techniques. The technical details of the implementation of atomistic simulations for the free-energy calculations are given in Sec.~\ref{sec:simulations}.

    The calculation of $\gamma_\t{gb}$ using the method described above [i.e., using Eq.~\eqref{eq:gamma_gb}] requires the free-energy calculations for at least the three different systems shown in Figs.~\ref{fig:simulation_boxes}(a)--(c). In addition to that, a fourth simulation system is required if the GB is asymmetric. This happens because in the case of asymmetric GBs the upper and lower surfaces in Fig.~\ref{fig:simulation_boxes}(c) will have different crystallographic orientations, thus two simulation systems of the type shown in Fig.~\ref{fig:simulation_boxes}(b) are required, one for each surface in the bicrystal system. In the remainder of this section we present an alternative approach that enables the calculation of $\gamma_\t{gb}$ for any type of GB (symmetric and asymmetric) using only two systems: one containing the GB, [Fig.~\ref{fig:simulation_boxes}(c)] and another composed of a perfect bulk [Fig.~\ref{fig:simulation_boxes}(a)].

    Consider a region of the bicrystal containing a GB shown in Fig.~\ref{fig:simulation_boxes}(d) by the blue dashed line. The slab has height $h$ and the same cross-sectional area as the GB. The thickness $h$ is chosen such that the upper and lower boundaries of the subsystem region are located inside the homogeneous bulk crystals. Accessing the thermodynamic properties of the subsystem allows the calculation of the GB free energy as \cite{Gibbs}:
    \begin{equation}
      \label{eq:gamma_gb_sub}
      \gamma_\t{gb} A_\t{gb} = F_\t{sub} - N_\t{sub} f_\t{bulk},
    \end{equation}
    where $F_\t{sub}$ is the total free energy of the subsystem, and $N_\t{sub}$ is the number of atoms contained in the subsystem. Thus, the subsystem defined in Fig.~\ref{fig:simulation_boxes}(d) circumvents the need to compute the surface free energy $\gamma_\t{surf}$: by computing $f_\t{bulk}$ using the bulk system of Fig.~\ref{fig:simulation_boxes}(a) and the free energy $F_\t{sub}$ of the bicrystal subsystem [Fig.~\ref{fig:simulation_boxes}(d)] one can obtain the GB free energy through Eq.~\eqref{eq:gamma_gb_sub}.

    In order to differentiate the two GB free energy calculation methods, we refer to the use of Eq.~\eqref{eq:gamma_gb_sub} as the ``subsystem approach'', while using Eq.~\eqref{eq:gamma_gb} will be referred to as the ``full system approach''. In Sec.~\ref{sec:results} we employ both approaches and demonstrate that they give consistent results.

  \subsection{Nonequilibrium Frenkel-Ladd method}
    The FL method \cite{fl} is a type of thermodynamic integration used to compute the absolute free energies, including all anharmonic effects, of simple crystalline systems from atomistic simulations such as MD or Monte Carlo.

    Consider the classical Hamiltonian of a system composed of $N$ interacting particles:
    \begin{equation}
      \label{eq:H_0}
      \mathcal{H}_0 = \sum_{i=1}^N \f{\v{p}_i^2}{2m} + U(\v{r}_1, \v{r}_2, \ldots \v{r}_N),
    \end{equation}
    where $\v{p}_i$ is the momentum of the $i$th particle, $m$ is the mass of the particles (we assume all particles to be identical), $\v{r}_i$ is the coordinate of the $i$th particle, and $U(\v{r}_1, \v{r}_2, \ldots \v{r}_N)$ is the many-body potential through which these particles interact with each other. We assume here that at the temperature and volume of interest this system is stable (or metastable) in a solid phase of known crystal structure. Considering this crystal structure we can construct the following Hamiltonian:
    \begin{equation}
      \label{eq:H_E}
      \mathcal{H}_\t{E} = \sum_{i=1}^N \f{\v{p}_i^2}{2m} + \sum_{i=1}^N \f{1}{2} m \omega^2 (\v{r}_i - \v{r}_i^0)^2,
    \end{equation}
    where each particle is attached to an equilibrium position $\v{r}_i^0$ by a harmonic spring with spring constant $k \equiv m \omega^2$. The set of equilibrium positions $\{\v{r}_i^0\}$ corresponds to the equilibrium crystal lattice positions of the particles in the system given by $\mathcal{H}_0$. This harmonic system is often called an Einstein crystal. Using Eqs.~\eqref{eq:H_0} and \eqref{eq:H_E}, the following Hamiltonian can be constructed:
    \begin{equation}
      \label{eq:H}
      \mathcal{H}(\lambda) = (1-\lambda) \mathcal{H}_0 + \lambda \mathcal{H}_\t{E},
    \end{equation}
    where $\lambda$ is a parameter. Notice that $\mathcal{H}(\lambda = 0) = \mathcal{H}_0$ and $\mathcal{H}(\lambda = 1) = \mathcal{H}_\t{E}$, hence, $\mathcal{H}(\lambda)$ is an interpolation between $\mathcal{H}_0$ and $\mathcal{H}_\t{E}$. The free energy of the system given by Eq.~\eqref{eq:H} is
    \begin{equation}
      \label{eq:F}
      F(N, V, T; \lambda) = - \kB T \ln \l[ \f{1}{h^{3N}} \int \d{\v{r}}^N \d{\v{p}}^N \e^{-\mathcal{H}(\lambda)/\kB T} \r].
    \end{equation}
    Furthermore, it can be shown from direct derivation of Eq.~\eqref{eq:F} that
    \begin{equation}
      \label{eq:dF}
      \pd{F}{\lambda} = \Avg{\pd{\mathcal{H}}{\lambda}}_\lambda,
    \end{equation}
    where $\avg{\ldots}_\lambda$ is a canonical ensemble average taken using $\mathcal{H}(\lambda)$ with a specific value of parameter $\lambda$. Equation \eqref{eq:dF} can be integrated in $\lambda$ from zero to one, resulting in:
    \begin{equation}
      \label{eq:F_0}
      F_0(N, V, T) = F_\t{E}(N, V, T) + \int_0^1 \Avg{U-U_\t{E}}_\lambda \d{\lambda}, 
    \end{equation}
    where $F_0(N, V, T)$ is the free energy of the system given by Eq.~\eqref{eq:H_0}, $F_\t{E}(N, V, T) = 3 N \kB T \ln ( \hbar\omega / \kB T)$ is the free energy of the Einstein crystal system given by Eq.~\eqref{eq:H_E}, and $U_\t{E}$ is the potential energy of the Einstein crystal, i.e., the second term in the right-hand side of Eq.~\eqref{eq:H_E}.

    In the FL method, Eq.~\eqref{eq:F_0} is used to compute the free energy $F_0(N, V, T)$ of the system of interacting particles given by $\mathcal{H}_0$ in Eq.~\eqref{eq:H_0}. The term $\avg{U-U_\t{E}}_\lambda$ is computed using atomistic simulations for $\lambda$ values ranging from zero to one, and the integral in the right-hand side of Eq.~\eqref{eq:F_0} is computed numerically. Since its first appearance the FL method has become increasingly efficient due to advances in the technique of thermodynamic integration. Most notably, the advent of nonequilibrium thermodynamic integration (also known as adiabatic switching \cite{as}) has increased the accuracy of the FL method substantially. In this paper, we follow closely the nonequilibrium FL method implementation described in Ref.~\cite{freitas_fl}. This implementation and the practices discussed in Ref.~\cite{freitas_fl} have been successfully applied to a variety of systems, such as calculation of the free energy of surface steps \cite{freitas2017step}, the study of structural phase transitions \cite{ma2017dynamic}, and the determination of melting temperatures \cite{melting}.

  \subsection{Quasi-harmonic approximation}
    In order to compute the free energy of the system described by $\mathcal{H}_0$ [Eq.~\eqref{eq:H_0}] in the QHA \cite{dove} we perform a second-order Taylor expansion of Eq.~\eqref{eq:H_0} around the equilibrium positions in the crystalline lattice at temperature $T$: $\v{r}^0(T) \equiv \{ \v{r}_1^0(T), \v{r}_2^0(T), \ldots, \v{r}_N^0(T)\}$, resulting in:
    \begin{align}
      \label{eq:taylor}
      \mathcal{H}_0 \approx \sum_{i=1}^N \f{\v{p}_i^2}{2m} & + \, U\big(\v{r}^0(T)\big) \\
       + \sum_{i=1}^N \sum_{j=1}^N & \sum_{\alpha,\beta} \f{m D_{i,j}^{\alpha,\beta}(T)}{2} (r_{i,\alpha} - r_{i, \alpha}^0) (r_{j,\beta} - r_{j, \beta}^0) \nonumber
    \end{align}
    where $r_{i,\alpha}$ is component $\alpha$ of the position of the $i$th particle $\v{r}_i$, where $\alpha = x$, $y$, or $z$, and 
    \[
      D_{i,j}^{\alpha,\beta}(T) \equiv \f{1}{m} \l( \f{\partial^2 U}{\partial r_{i,\alpha} \partial r_{j,\beta}} \r)_{\v{r} = \v{r}^0(T)}
    \]
    are the components of the potential energy Hessian matrix. The equations of motion of this harmonic system can be uncoupled by performing a canonical transformation:
    \begin{equation}
      \label{eq:taylor_2}
      \mathcal{H}_0 \approx \sum_{n=1}^{3N} \l[ \f{\tilde{p}_n^2}{2m} + \f{1}{2} m \Omega_n^2(T) \tilde{q}_n^2 \r],
    \end{equation}
    where $\Omega_n(T)$ are the eigenvalues of $\v{D}(T)$. The Taylor-expanded Hamiltonian in Eq.~\eqref{eq:taylor_2} is a quadratic system, hence its free energy can be computed analytically:
    \begin{equation}
      \label{eq:F_ha}
      F_\t{qha}(N, V, T) \approx \kB T \sum_{n=1}^{3(N-1)} \ln \fp{\hbar \Omega_n(T)}{\kB T},
    \end{equation}
    where we removed from the free energy expression the three null eigenvalues of $\v{D}$. Equation \eqref{eq:F_ha} is the free energy of system $\mathcal{H}_0$, Eq.~\eqref{eq:H_0}, in the QHA.

    Notice that in the QHA the Taylor expansion, Eq.~\eqref{eq:taylor}, is performed around the equilibrium lattice positions at finite temperature $\v{r}^0(T)$. This is in contrast with the harmonic approximation, where the Taylor expansion is performed around $\v{r}^0(T=0)$, i.e., the energy minimum of the potential energy surface. Thus, the QHA improves on the harmonic approximation by incorporating anharmonic effects due to the thermal expansion of the solid. In practice, the QHA method requires the phonon frequencies $\Omega_n$ to be recomputed for each temperature $T$ after expanding the system's volume to account for thermal expansion.

\section{Atomistic simulations\label{sec:simulations}}
  \subsection{Molecular Dynamics simulations}
    We have employed MD simulations using the Large-scale Atomic/Molecular Massively Parallel Simulator (LAMMPS) \cite{lammps} software. The interactions between the atoms were modeled using the embedded-atom method (EAM) \cite{eam} potential for Cu from \citet{mishin_cu}. The timestep of the MD simulations was $2\,\t{fs}$, which is equivalent of $1/64\,\t{th}$ of the period of oscillation of the highest frequency phonon of this system. In order for the MD simulations to sample the canonical ensemble distribution we applied the Langevin \cite{langevin,allen_tildesley} thermostat with a relaxation time of $\tau \equiv m / \gamma = 2\,\t{ps}$, where $\gamma$ is the friction parameter of the thermostat and $m = 63.546\,\t{g/mol}$ is the Cu atomic mass.

    The nonequilibrium FL simulations were performed with a switching time of $0.9\,\t{ns}$ with $0.1\,\t{ns}$ of equilibration before the forward switching and between the forward and backward switching. The $\lambda(t)$ parameter followed the $S$-shaped functional form \cite{fl_noneq}, while the spring constant for the Einstein crystal, Eq.~\eqref{eq:H_E}, was obtained from the mean-squared displacement of the atoms in a perfect bulk system. For all free-energy values reported in this paper, the estimates of the measurement error were obtained by performing ten independent FL simulations.

  \subsection{System geometry\label{sec:geometry}}
\begin{figure}
      \centering
      \includegraphics[width=0.48\textwidth]{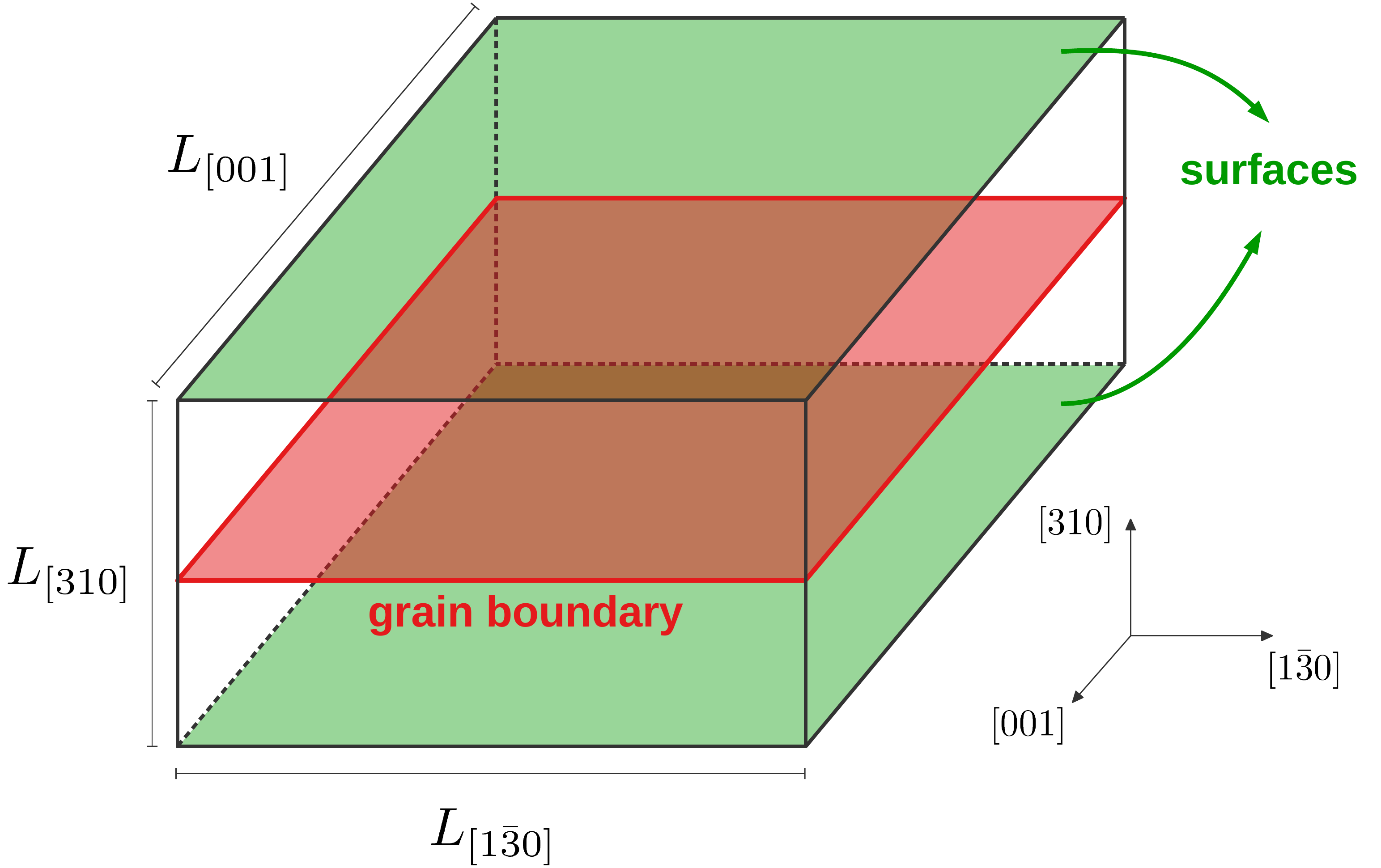}
      \caption{\label{fig:geometry}Bicrystal system geometry, equivalent to Fig.~\ref{fig:simulation_boxes}(c). The crystallographic orientation is such that both free surfaces are $\{310\}$ and the GB is a $\Sigma 5 (310)[001]$.}
    \end{figure}

    The bicrystal system geometry is shown in Fig.~\ref{fig:geometry} along with the crystallographic orientations for the $\Sigma 5 (310) [001]$ GB. Periodic boundary conditions were applied in the directions parallel to the GB plane, while free surface boundary conditions were applied in the $[310]$ direction, resulting in two $\{310\}$ surfaces. The simulation system height $L_{[310]} = 71.0 \Ang$ was chosen large enough to make the effect of the interaction between the elastic fields of the GB and surface negligible. The determination of the system dimensions parallel to the GB require accounting for finite-size effects due to thermal motion of the atoms, therefore, we postpone the analysis of the size convergence of the results with $A_\t{gb} = L_{[1\bar{3}0]} \times L_{[001]}$ until Sec.~\ref{sec:results} where the GB free energy convergence with $A_\t{gb}$ is measured directly. For the sake of completeness we quote here the converged dimensions as concluded from the analysis of Sec.~\ref{sec:results}: $A_\t{gb} = 44.7 \, \t{nm}^2$, with $L_{[001]} = 68.6 \Ang$ and $L_{[1\bar{3}0]} = 65.1 \Ang$. All system dimensions have been expanded at finite temperatures to accommodate the thermal expansion of the crystal. The total number of atoms of the bicrystal system depends on the GB phase because of their different GB densities, for the NK boundary we have $N_\t{nk} = 27,000$ atoms and for the SK boundary we have $N_\t{sk} = 27,096$ atoms.

    The calculation of the GB free energy using the full system approach requires the simulation of at least two other systems aside from the bicrystal in Fig.~\ref{fig:geometry}. The second system consists of a single-crystal slab with two $(310)$ surfaces: this system has identical dimensions, boundary conditions, and crystallographic orientation as the bicrystal but it does not contain the GB in the center of the system, as shown in Fig.~\ref{fig:simulation_boxes}(b). This system contained $N_\t{surf} = 27,000$ atoms. The third simulation system employed is a perfect bulk in cubic shape where full periodic boundary conditions were applied. This system, with $N_\t{bulk} = 27,436$ atoms, has the simulation box edges aligned with the $\avg{100}$ crystallographic directions as shown in Fig.~\ref{fig:simulation_boxes}(a). For the subsystem approach, in addition to the bicrystal system in Fig.~\ref{fig:geometry}, we only use the single-crystal slab system described above.

\section{Results\label{sec:results}}
  \subsection{Convergence with the GB area}
    The grain boundary free energy $\gamma_\t{gb}$ has size dependence at small GB areas due to in-plane phonon modes along the GB. Hence, a set of simulations was performed to identify the smallest GB area, $A_\t{gb}$, that gives size-independent results for the GB free energy. The system height, $L_{[310]}$, was kept constant while simulations were carried out for different system cross sections $L_{[001]} \times L_{[1\bar{3}0]}$, where the cross section was kept as close to a square as possible.

\begin{figure}[hbt]
      \centering
      \includegraphics[width=0.48\textwidth]{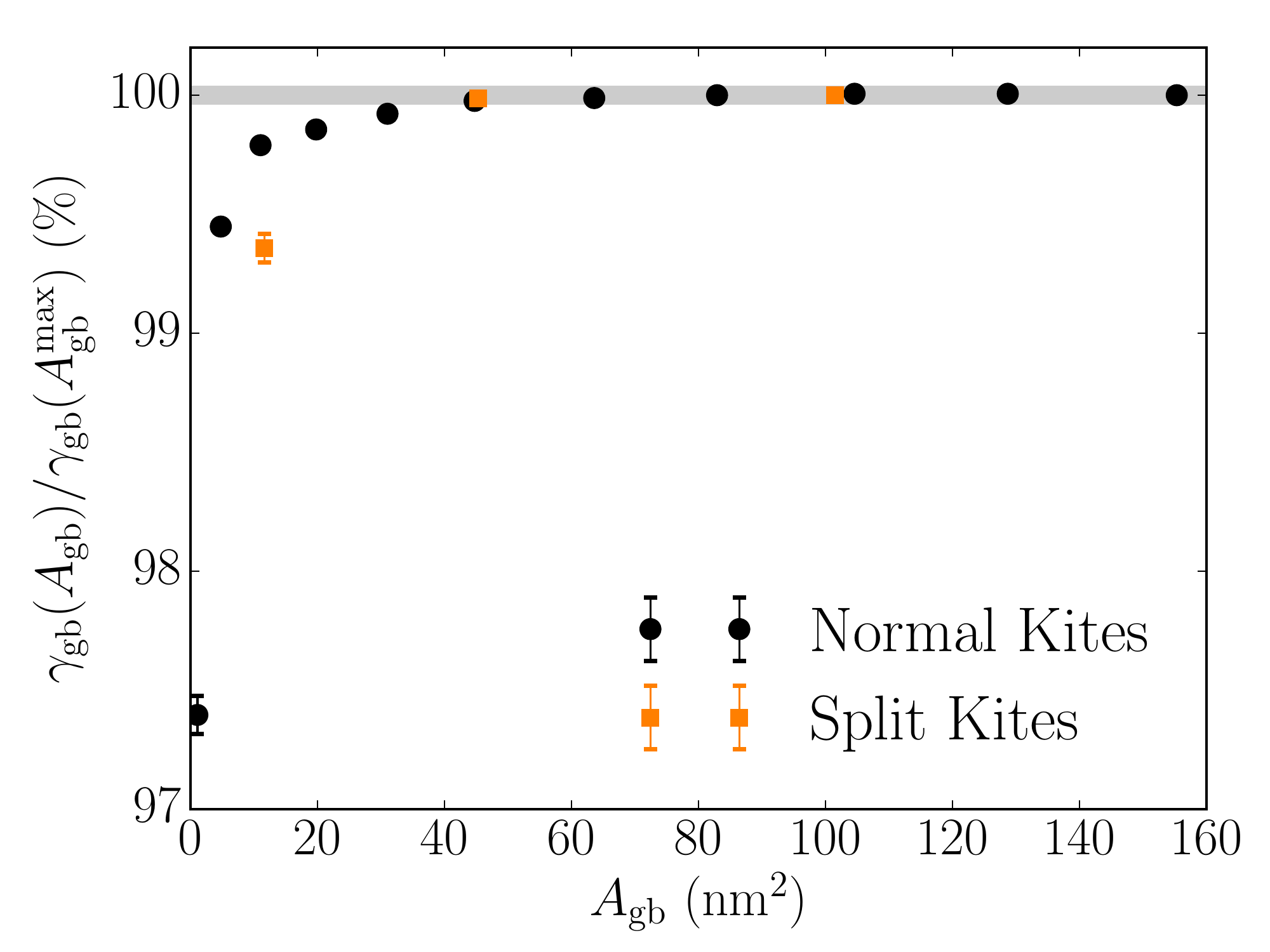}
      \caption{\label{fig:A_convergence}Convergence of the GB free energy at $T = 200\K$ with the GB area. The result is shown as the ratio of $\gamma_\t{gb}$ at the area $A_\t{gb}$ to its value at the largest area considered: $A^\t{max}_\t{nk} = 155.5 \, \t{nm}^2$ and $A^\t{max}_\t{sk} = 101.5 \, \t{nm}^2$. The horizontal gray line width represents the error bar in $\gamma_\t{gb}$ at the largest considered GB area. Based on this figure, we have employed $A_\t{gb} = 44.7\,\t{nm}^2$ for the remaining free-energy calculations.}
    \end{figure}

    Figure \ref{fig:A_convergence} shows the convergence of the GB free energy with the GB total area for $T = 200\K$. The convergence of the NK phase free energy is fast: the ratio of $\gamma_\t{nk}(A_\t{nk})$ for the smallest size considered, $1.25\,\t{nm}^2$, to the largest size considered, $A_\t{nk}^\t{max} = 155.5\,\t{nm}^2$, is $(97.4 \pm 0.1)\%$ despite the two orders of magnitude difference in GB area. This ratio is increased to $(99.79 \pm 0.04)\%$ if the GB area is increased to $11.2\,\t{nm}^2$, while for $A_\t{nk} > 20\,\t{nm}^2$ the ratio becomes equal to one within the error bar. For the SK phase the data granularity is much coarser because the smallest unit is a 10$\times$2 reconstruction relative to NK, nevertheless we still see that for $A_\t{sk} > 20\,\t{nm}^2$ the SK phase free energy ratio with respect to the largest area considered is equal to one within the error bar. Hence, for the remainder of this paper we have adopted $A_\t{gb} = 44.7 \, \t{nm}^2$ for the free-energy calculations of both GB phases.

  \subsection{Comparison of full system and subsystem approaches}
    The subsystem approach for GB free energy calculation requires the subsystem boundaries to be located far from the influence of both: the free surface and the GB itself. In order to verify the convergence of the subsystem approach with the subsystem height $h$ we compare the ratio of the GB free energy computed using the subsystem approach with the free energy as obtained from the full system approach. For these free-energy calculations the system dimensions were kept constant while the subsystem height $h$ was varied. 

\begin{figure}[hbt]
      \centering
      \includegraphics[width=0.48\textwidth]{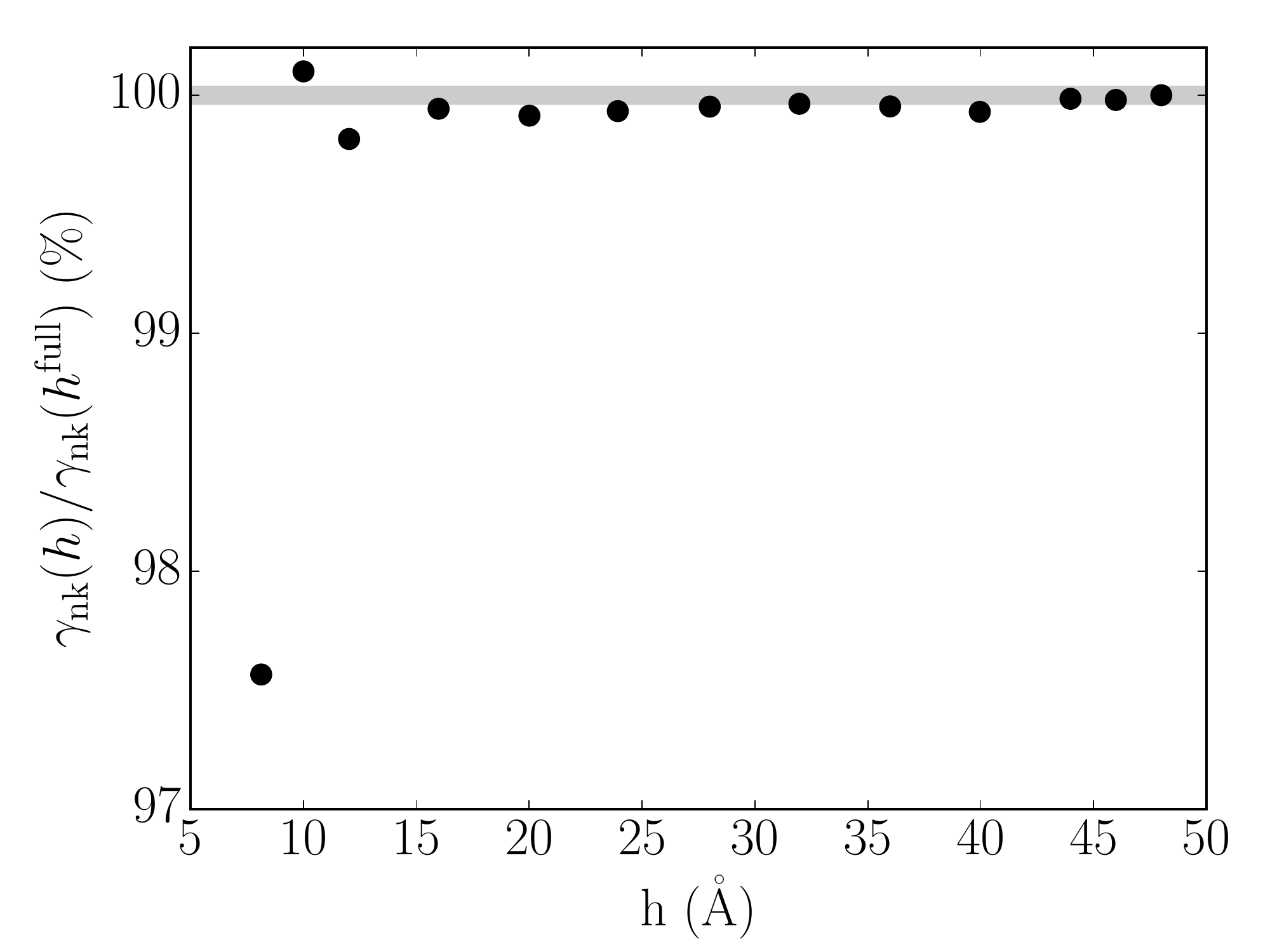}
      \caption{\label{fig:h_convergence}Convergence of the NK phase free energy at $T = 200\K$ as computed using the subsystem approach with the subsystem height $h$, shown in Fig.~\ref{fig:simulation_boxes}(d). The result is normalized by the free energy of the NK phase as computed using the full system approach. The horizontal gray line width represents the error bar in the free energy as computed from the full system approach. Based on this figure we have employed $h = 35.5 \Ang$ for the following remaining free-energy calculations.}
    \end{figure}

    Figure \ref{fig:h_convergence} shows the convergence of the NK boundary free energy with the subsystem height $h$ for $T = 200\K$. The GB free energy is recovered within $2.5\%$ by considering only atoms within a subsystem with $h = 8\Ang$ (i.e., $11\%$ of the system's atoms) and within $0.2\%$ for atoms within $h = 12\Ang$ subsystem (i.e., $17\%$ of the system's atoms). Therefore, the GB free energy computed using the subsystem approach is consistent with the full system approach. In the next section we show that this approach is also more accurate and results in smaller error bars than the full system approach.

    The agreement of the full system and the subsystem approaches also demonstrates the convergence of the free-energy calculations with respect to the slab thickness $L_{[310]}$. In Fig.~\ref{fig:h_convergence} we observe that for the subsystem height in the range $15 \Ang \le h \le 45\Ang$ the GB free energy reaches a plateau and agrees with the full system approach within the error bar, indicating convergence with respect to the slab thickness. Were the calculations not converged with respect to the system thickness, the elastic fields of the GB and the surfaces would interact, thus there would be no range of $h$ for which Eqs.~\eqref{eq:gamma_gb} and \eqref{eq:gamma_gb_sub} give the same free energy. Notice how in the nonconverged case $F_\t{gb}$ in Eq.~\eqref{eq:gamma_gb} would capture the entirety of this elastic interaction, while $F_\t{sub}$ in Eq.~\eqref{eq:gamma_gb_sub} would continually capture more of this elastic interaction as $h$ increases, effectively capturing all of the elastic interaction only at $h = L_{[310]}$, at which point Eqs.~\eqref{eq:gamma_gb} and \eqref{eq:gamma_gb_sub} result in different GB free energies because of the lack of the surface free-energy term in Eq.~\eqref{eq:gamma_gb_sub}.

    For the remainder of the paper, we have put the subsystem boundaries halfway between the GB and free surfaces, resulting in $h = L_{[310]}/2 \approx 35.5 \Ang$.

  \subsection{GB phase transition}
\begin{figure}[hbt]
      \centering
      \includegraphics[width=0.48\textwidth]{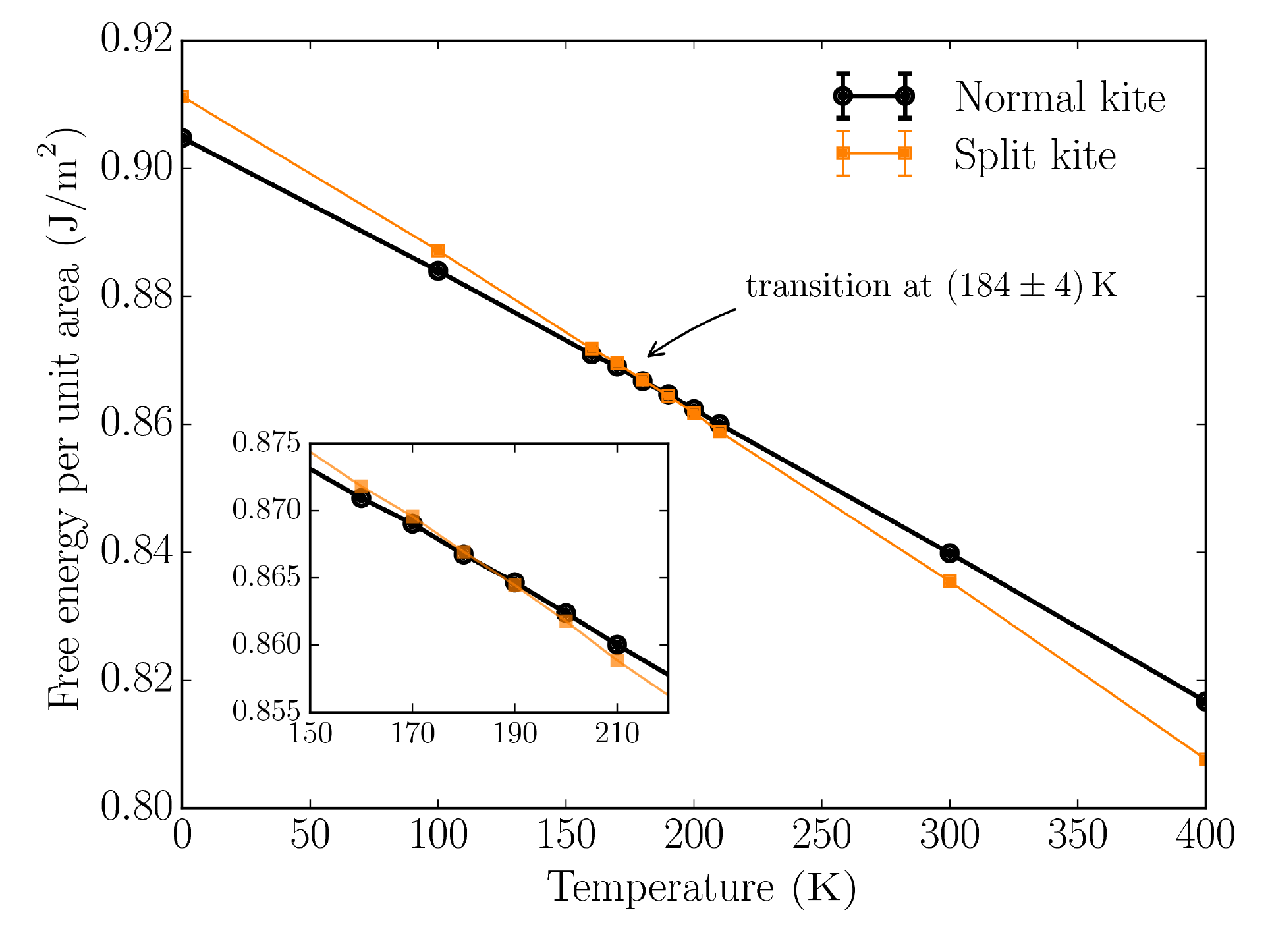}
      \caption{\label{fig:transition}Temperature dependence of the free energy of NK and SK phases as computed with the FL method using the subsystem approach. The crossing of the curves at $T^* = (184 \pm 4) \K$ is an evidence that there is a well defined thermodynamic phase transition between the two structures. The phase ordering observed is consistent with previous MD simulations where open surfaces acted as source and sinks of atoms to induce the transition \cite{Frolov2013}.}
    \end{figure}

    Figure \ref{fig:transition} shows the free energies of NK and SK phases as a function of temperature, as computed with the FL method using the subsystem approach. Consistent with previous calculations \cite{Frolov2012b,Foiles2010,obrien,Foiles94,Broughton86a} the free energy decreases by about $10\%$ in the temperature interval from $0\K$ to $400\K$. The slope of the curves are different and the free energy lines cross at $T^* = (184 \pm 4)\K$, predicting the temperature of the thermally induced GB phase transition. The crossover temperature was calculated by fitting the free energy data points in the interval from $160\K$ to $210\K$ using linear regression, the error bar was estimated from the fitting procedure by taking into consideration the data points error bars. Below $T^*$ the NK is stable, while at high temperatures the SK phase becomes more stable due to its higher entropic contribution to the free energy. At $T^*$ the two GB phases are expected to coexist in equilibrium in an open system connected to sources and sinks of atoms. The free-energy calculations are consistent with the observation in MD simulations with open surfaces that the SK structure, while higher in energy than NK at $T=0\K$, is the most stable state at $T \ge 800\K$ \cite{Frolov2013}.

\begin{figure}[hbt]
      \centering
      \includegraphics[width=0.48\textwidth]{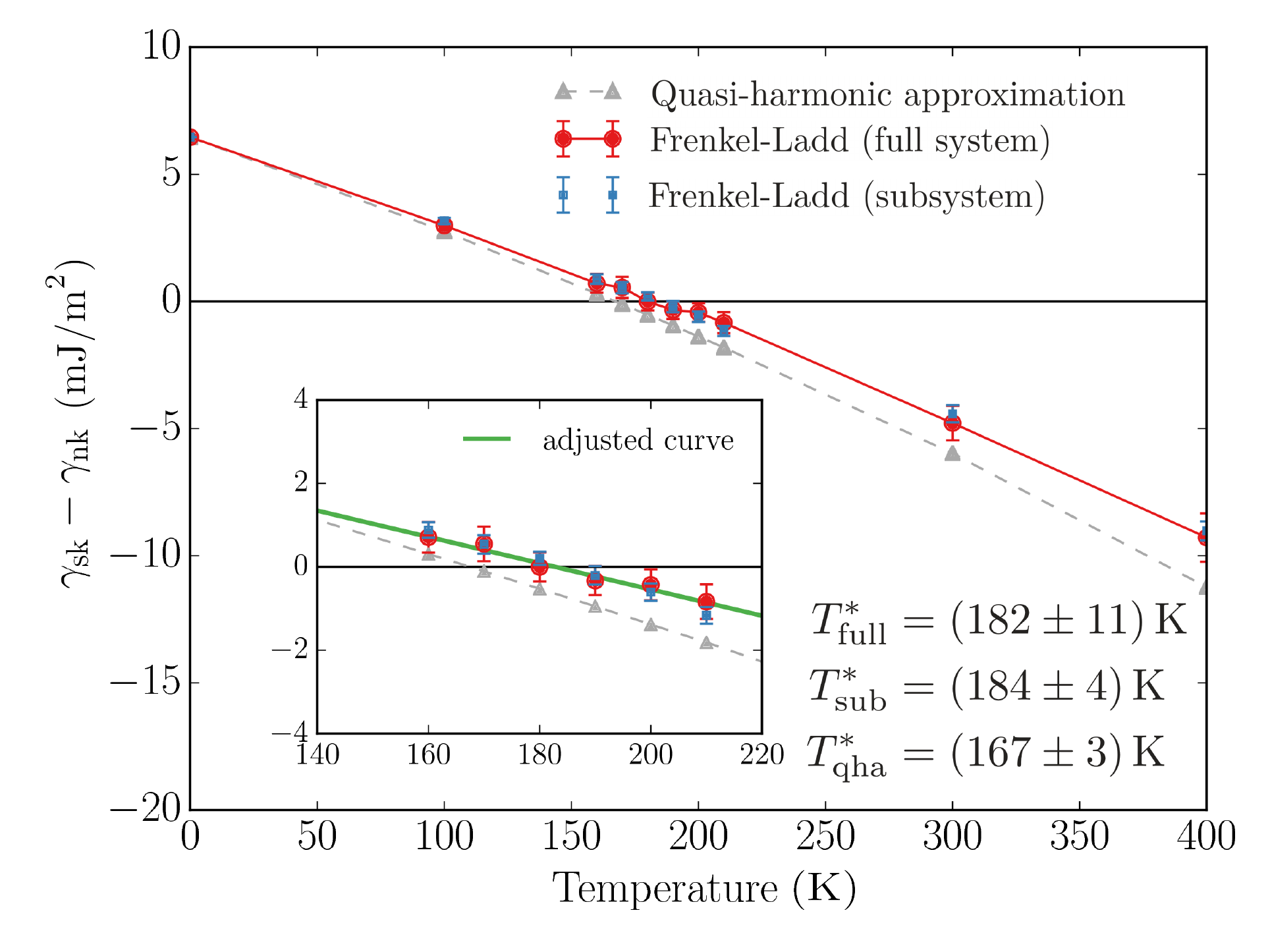}
      \caption{\label{fig:difference}Difference in free energy between the NK and the SK phases as computed using different methods. Both approaches using the FL method (full system and subsystem approaches) result in the same transition temperature within the error bars. The QHA consistently identifies the SK as the high-temperature GB phase, although the transition temperature is $9\%$ lower than the FL method prediction.}
    \end{figure}

    The results of the free-energy calculations using the FL method in the full system and subsystem approaches are compared in Fig.~\ref{fig:difference}, where the free energy difference $\gamma_\t{sk}-\gamma_\t{nk}$ is shown as function of temperature. We observe that both approaches predict transition temperatures that agree within the error bars: $T^*_\t{sub} = (184\pm4)\K$ and $T^*_\t{full} = (182\pm 11)\K$, however the subsystem approach results in a smaller error bar at all temperatures. This happens because during the nonequilibrium thermodynamic integration switching the subsystem approach avoids the dissipation that occurs due to the anharmonicity of the surface region atoms. This improves the accuracy of the results in two ways, first the systematic error due to the dissipation present in the switching is decreased due to lower total dissipation. Second, the magnitude of the fluctuations in $\avg{U-U_\t{E}}$, Eq.~\eqref{eq:F_0}, also decreases due to the diminished anharmonicity of the system, resulting in a smaller magnitude of the random error.

    Figure \ref{fig:difference} also presents the GB phases free energy difference as computed from QHA method, shown as the gray triangle data points. The QHA calculations are relatively inexpensive, however they make approximations about the nature of the atomic vibrations. Indeed, a previous systematic comparison of several different GB free energy calculation methods indicated that the QHA becomes less accurate at higher temperatures \cite{Foiles94}. Notice that the FL method makes no assumptions about the atomic vibrations and, thus, it is expected to be accurate even at relatively high temperatures. Comparing the QHA with the FL method in Fig.~\ref{fig:difference} we observe that the QHA systematically underestimates the free energy difference: it predicts a transition temperature of $T^*_\t{qha}= (167 \pm 3) \K$, which is $9\%$ lower than $T^*_\t{sub}$. This discrepancy suggests that while the anharmonic effects may be considered negligible for the absolute values of the GB free energies at low temperatures \cite{Foiles94}, they produce non-negligible changes in the calculated transition temperature. It is interesting to notice, however, that the QHA correctly identify the phase ordering. At this point it is important to notice that both methods, FL and QHA, are not applicable in the presence of configurational disorder because both assume that atomic motion happen around each atoms' equilibrium position. In the system studied here we did not observe any atomic diffusion along the GBs for the temperature range considered.

    Figures \ref{fig:A_convergence} and \ref{fig:h_convergence} raise the question of what is the smallest subsystem containing the GB that can use be used to compute $\gamma_\t{gb}$ fully converged with respect to system dimensions. These figures show that for $A_\t{gb} = 5.0 \, \t{nm}^2$ and $h = 12\Ang$ the free energy is recovered within $0.7\%$ when compared to the free energy of system with $A_\t{gb}$ two orders of magnitude larger and $h = L_{[1\bar{3}0]}$ (i.e., using the full system approach). We conclude that atoms contributing to the GBs free energy, and thus to the phase transition driving force, are contained in a small subsystem volume with $500$ atoms only. This shows that the changes in the vibrational properties of the material that account for the phase transformation are strongly localized around the GB (due to the low value for $h$) and are not due to long wavelength phonons along the GB (because of the small value for $A_\t{gb}$). These values are likely to vary with the GB character.

\begin{figure}[hbt]
      \centering
      \includegraphics[width=0.48\textwidth]{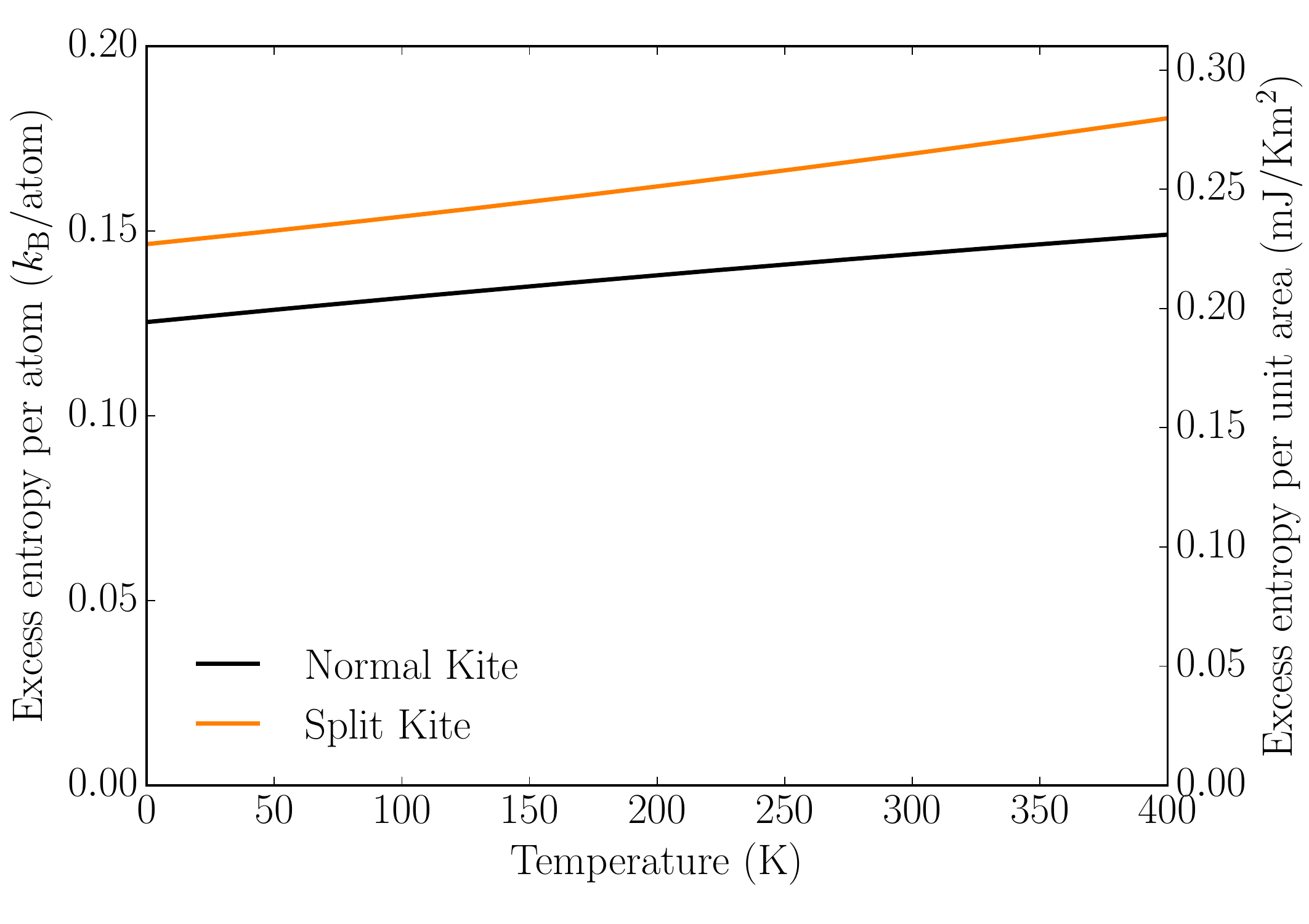}
      \caption{\label{fig:entropy}Average excess entropy per atom due to the GB, determined by distributing the excess entropy per unit area equally between all atoms within $h = 12\Ang$ from the GB plane. These values are between $10\%$ and $15\%$ of the bulk entropy per atom.}
    \end{figure}

    From the temperature dependence of the free energy curves in Fig.~\ref{fig:transition} it is possible to estimate the excess entropy of both GB phases. The temperature dependence of each GB phase is described by the adsorption equation \cite{Gibbs,Cahn79,Frolov09a,Frolov2012a} 
    \begin{equation}
      \label{eq:adsorption}
      \d{\gamma_\t{gb}} = - [s]_N \d{T} + \sum_{i,j=1,2} (\tau^\t{gb}_{ij}-\gamma_\t{gb}) \d{\e_{ij}}
    \end{equation}
    where $[s]_N$ is the excess GB entropy per unit area, $\tau_\t{gb}$ is the GB stress, and $\e_{ij}$ is the elastic strain tensor. Neglecting the small contribution of $\tau$ \cite{Frolov2012b}, which is equal to the GB mechanical work against thermal expansion, we can calculate the excess entropy $[s]_N$ of each GB phase as minus the derivative of the free energy with respect to the temperature. Because we have estimated above that only atoms within a small $h = 12\Ang$ subsystem contribute to the GB free energy we can convert the entropy per unit area to average entropy per atom, which is a well documented quantity for atoms in the bulk. Figure \ref{fig:entropy} shows the average excess entropy per atom for both GB phases as a function of temperature. From this figure we see that SK has a higher entropy per atom than the NK, but both excess entropies are on the order of $0.1 \, \kB / \t{atom}$. Thus, the excess entropy per atom for these phases is between $10\%$ and $15\%$ the entropy per atom of the bulk phase. For both structures the entropy dependence on the temperature is almost linear, with a small slope of about $\approx 7 \times 10^{-5} \, \kB/(\t{atom} \, . \K)$. We notice that the average excess entropy per atom computed here is of the same order of magnitude as the excess entropy per atom reported based on inelastic neutron scattering experiments in nanocrystalline Fe and Ni$_3$Fe: $0.1$ to $0.4\, \kB / \t{atom}$ \cite{exp_1,exp_2,exp_4}.

\section{Discussion and Conclusions\label{sec:conclusions}}
  In this work, we have studied GB phase transitions using the free-energy calculations. The approach is computationally efficient and accurately predicts free energies of different GB phases in the temperature range below about half of the melting point. This range is particularly important since atomistic simulations are not effective at modeling GB phase transitions at these low temperatures. The GB free energies were calculated using FL and QHA methods. The FL method was shown to be more accurate than the QHA because it makes no assumptions about the nature of the GB atom vibrations, as demonstrated in previous studies \cite{Foiles94}.

  The inherent multiplicity of GB structures often makes it challenging to predict the finite-temperature structures based on the results of $0\K$ calculations alone. For example, recent calculations demonstrated that SK and NK structures of different $[100]$ symmetric tilt boundaries in copper have close energies. While the energy difference at $0\K$ was only few percent and was found to be both positive and negative for different GBs, all these boundaries transformed from NK to SK at high temperature \cite{zhu2018}. Similar multiplicity of structures and preferred high-temperature GB phases have been demonstrated for $[110]$ symmetric tilt GBs in bcc tungsten \cite{frolov2018grain,frolov2018structures}. The apparent high-temperature stability of certain types of GB structures in these systems is not well understood. Both vibrational and configurational entropy are expected to contribute to the GB free energy. In this study we demonstrated that the vibrational entropy stabilizes the SK GB phase of the $\Sigma5 (310) [001]$ GB even at relatively low temperature. The computed value for entropy per atom for both phases are on the order of $0.1\kB/\t{atom}$, equivalent to about $15\%$ of the entropy per atom of the bulk phase. These values are on the same order of magnitude as the values obtained from neutron scattering experiments \cite{exp_1,exp_2,exp_4} for nanocrystalline Fe and Ni$_3$Fe.

  Using a model system of a $\Sigma5 (310) [001]$ GB in Cu we calculated free energies of NK and SK phases. The two structures correspond to the two minima on the curve of GB energy as a function of GB density. The NK phase is the ground state at low temperatures, but at $T^* = (184\pm4)\K$ the free-energy curves cross, suggesting a transition to the SK phase. At higher temperatures, the SK phase has lower free energy and, thus, becomes more stable than NK. The free-energy analysis is consistent with previous MD simulations that demonstrated SK to be the most stable state at $T \ge 800\K$. Using this model system we demonstrated that $0\K$ GB structure search complemented with free-energy calculations can predict finite-temperature GB structure and characterize GB phase diagrams in a model elemental system. 

  The transition temperate of $184\K$ calculated for our model system is much lower than the transition temperate suggested by the diffusivity measurements in the same boundary. In the experiments, the change in the Arrhenius slope of the diffusion flux occurred in the temperature interval from $700\K$ to $900\K$ for Ag \cite{Divinski2012} and even slightly higher for Au \cite{au}. The $0\K$ energy difference between the two GB phases is very small and taking into account the strong temperature dependence of the GB free energy, such discrepancies between the model predictions and the experimental measurements could be expected. Nevertheless, the calculations clearly identify SK as a preferable high-temperature structure due to its entropic properties. Finally, we note that the diffusivity measurements were performed for systems with impurities. According to calculations for the $\Sigma 5(210)[001]$ and $\Sigma5(310)[001]$ boundaries in Cu, Ag atoms segregate to NK structure much stronger that to SK due to its more open structure \cite{PhysRevB.92.020103,Frolov2013PRL}. As a result, it is expected that addition of Ag would stabilize NK and raise the transitions temperature, bringing modeling and experiments into closer agreement.

  Although the FL method provides a precise framework for computing vibrational contributions to the free energy at any arbitrarily high temperature, the presence of appreciable configurational disorder established through GB diffusion at high homologous temperatures would prohibit its application \cite{freitas_fl}. The free-energy curves calculated in this work can be extended to higher temperatures in a straightforward manner by integrating the adsorption equation \cite{Frolov2012a,Frolov2012b} [Eq.~\eqref{eq:adsorption}]. This approach has been demonstrated to predict GB free energy as a function of misorientation for $\Sigma3$ boundaries \cite{obrien}. Although GB phase transitions at high temperature can be simulated directly, the free energy provides useful information about the magnitude of the driving force for such transformations. The GB free energy in an elemental system can also serve as a reference value for calculations in a binary or more general multicomponent system. 

  \begin{acknowledgments}
    R.F. acknowledges support from the Livermore Graduate Scholar Program. This work was performed in part under the auspices of the U.S. Department of Energy by Lawrence Livermore National Laboratory under Contract No. DE-AC52-07NA27344. The work was supported by the Laboratory Directed Research and Development Program at LLNL, project 17-LW-012. M.A. acknowledges support from the US National Science Foundation under grant No. DMR-1507033.
  \end{acknowledgments}
  \bibliography{bibliography}
\end{document}